\documentclass[preprint]{revtex4}

\usepackage{bm}
\usepackage{amsmath,amssymb}
\usepackage{graphicx}
\usepackage{wrapfig}
\usepackage{color}

\newcommand{\p}{\partial}

\newcommand{\diff}{\mathrm{d}}

\newcommand{\nuii}{\nu_{\mathrm{ii}}}
\newcommand{\nuee}{\nu_{\mathrm{ee}}}

\newcommand{\nuei}{\nu_{\mathrm{ei}}}

\newcommand{\nue}{\nu_{\mathrm{e}}}

\newcommand{\mi}{m_{\mathrm{i}}}
\newcommand{\me}{m_{\mathrm{e}}}
\newcommand{\Ti}{T_{0\mathrm{i}}}
\newcommand{\Te}{T_{0\mathrm{e}}}
\newcommand{\rhoi}{\rho_{\mathrm{i}}}
\newcommand{\rhoe}{\rho_{\mathrm{e}}}
\newcommand{\rhos}{\rho_{\mathrm{Se}}}
\newcommand{\de}{d_{\mathrm{e}}}
\newcommand{\di}{d_{\mathrm{i}}}

\newcommand{\tauA}{\tau_{\mathrm{A}}}
\newcommand{\VA}{V_{\mathrm{A}}}
\newcommand{\betae}{\beta_{\mathrm{e}}}
\newcommand{\betai}{\beta_{\mathrm{i}}}

\newcommand{\revise}[1]{\textcolor{black}{#1}}
\newcommand{\referee}[1]{\textcolor{black}{#1}}

\begin{document}
\title{Electron and Ion Heating during Magnetic Reconnection in Weakly
Collisional Plasmas}

\author{Ryusuke Numata}
\email{ryusuke.numata@gmail.com}
\affiliation{Graduate School of Simulation Studies, University of Hyogo,
7-1-28 Minatojima Minami-machi, Chuo-ku, Kobe, Hyogo 650-0047, Japan}
\author{Nuno F. Loureiro}
\affiliation{Associa\c{c}\~ao EURATOM/IST, Instituto de Plasmas e Fus\~ao
Nuclear---Laborat\'orio Associado,
Instituto Superior T\'ecnico,
Universidade T\'ecnica de Lisboa, 1049-001 Lisboa, Portugal}
\begin{abstract}
Gyrokinetic simulations of magnetic reconnection are presented to
investigate plasma heating for strongly magnetized, weakly collisional
plasmas. For a low plasma beta case, parallel and perpendicular phase
mixing strongly enhance energy dissipation yielding 
electron heating. Heating occurs for a long time period after a
dynamical process of magnetic reconnection ended. For a higher beta
case, the ratio of ion to electron dissipation rate increases,
suggesting
\revise{that ion heating (via phase-mixing) may become an important
dissipation channel in high beta plasmas.}
\end{abstract}
\maketitle

\section{Introduction}

Magnetic reconnection is a commonly observed fundamental process in both
astrophysical and fusion plasmas. It allows topological change of
magnetic field lines, and converts the free energy stored in the
magnetic field into various forms of energy, such as bulk plasma
flows, heating of plasmas, or non-thermal energetic particles. One of
the key issues of magnetic reconnection is energy partition, yet it has
not been widely studied. Here, we focus on plasma heating---how much
free \revise{magnetic} energy is transformed into the thermal energy
during magnetic 
reconnection? To address thermodynamic properties of plasmas, 
inter-particle collisions are especially important even though plasmas
are \revise{weakly collisional in many} environments where
magnetic reconnection occurs.


In weakly collisional plasmas, various kinetic effects play crucial
roles. Landau damping~\cite{Landau_46} is one of the well known
\revise{examples} of such kinetic effects, in which nearly synchronous
particles with waves absorb energy from the waves. The phase  
mixing effect of the Landau damping process creates progressively
oscillatory structures in velocity space. 
Such small scale
structures suffer strong collisional dissipation as the collision
operator provides diffusion in velocity space. As a result, as long as
collisions are sufficiently infrequent, the rate of energy dissipation
stays finite and is independent of collision \revise{frequency~\cite{NumataLoureiro_12,LoureiroSchekochihinZocco_13}}.
Similar phase mixing is also induced by a finite Larmor radius (FLR)
effect~\cite{DorlandHammett_93}. In strongly magnetized plasmas,
particles undergo drifts \revise{(dominantly the $\bm{E}\times\bm{B}$ drift)} in
the perpendicular direction to the mean
magnetic field. Gyro-averaging of the fields will give rise to
spread of the drift velocity of different particles, hence will lead
\revise{to} phase mixing in the perpendicular 
direction. Unlike linear parallel phase mixing of Landau damping,
the FLR phase mixing process causes damping proportional to the field
strength, and only appears nonlinearly (nonlinear phase mixing).
Linear~\cite{WatanabeSugama_04a} and nonlinear phase
mixing~\cite{TatsunoDorlandSchekochihin_09} have been studied numerically
in electrostatic turbulence.
\revise{The} importance of the phase mixing mechanism for energy
dissipation in solar wind turbulence has been verified
in~\cite{HowesTenBargeDorland_11}. Numerical simulations of magnetic
reconnection using a hybrid fluid/kinetic
model~\cite{ZoccoSchekochihin_11} have shown strong electron heating via 
\revise{linear} Landau damping for low-$\beta$
plasmas (\revise{$\beta$ is the ratio of the plasma to the magnetic
pressure})~\cite{LoureiroSchekochihinZocco_13}.

In this paper, we present gyrokinetic simulations of magnetic
reconnection using {\tt AstroGK}~\cite{NumataHowesTatsuno_10}.
We follow~\cite{NumataDorlandHowes_11} to setup a tearing instability
\revise{configuration}, and extend it to the nonlinear regime.
\revise{Even though the collision frequencies we choose are sufficiently
small so that reconnection occurs in the collisionless regime, 
we show that significant collisional dissipation resulting in background plasma
heating occurs via phase-mixing.}
We also consider the heating ratio of
electrons and ions, and its dependence on plasma $\beta$.

\section{Problem setup}

We consider magnetic reconnection of strongly magnetized 
plasmas in a two-dimensional doubly periodic slab domain.
We initialize the system by a tearing unstable magnetic field
configuration (see~\cite{NumataHowesTatsuno_10,NumataDorlandHowes_11}
for details). The equilibrium magnetic field profile is
\begin{equation}
\bm B=B_{z0}\hat z+B_{y}^{\mathrm{eq}}(x)\hat y, \quad B_{z0}\gg
 B_{y}^{\mathrm{eq}},
\end{equation}
where $B_{z0}$ is the background guide magnetic field and
$B_{y}^{\mathrm{eq}}$ is the in-plane, reconnecting component, related
to the parallel vector potential by \revise{$B_{y}^{\mathrm{eq}}(x)= -\p
A_{\parallel}^{\mathrm{eq}}/\p x$}, and
\begin{equation}
 \label{equilib_apar}
  A_{\parallel}^{\mathrm{eq}}(x) =
  A_{\parallel0}^{\mathrm{eq}}\cosh^{-2}\left(\frac{x-L_{x}/2}{a}\right)
  S_{\mathrm{h}}(x).
\end{equation}
($S_{\mathrm{h}}(x)$ is a shape function to enforce periodicity~\cite{NumataHowesTatsuno_10}.)
$A_{\parallel}^{\mathrm{eq}}$ is generated by the electron parallel
current to satisfy the parallel Amp\`ere's law.
The equilibrium scale length is denoted by $a$ and $L_x$ is the length
of the simulation box in the $x$ direction, set to $L_x/a=3.2\pi$. In
the $y$ direction, the box size is $L_y/a=2.5\pi$. We impose a small
sinusoidal perturbation to the equilibrium magnetic field,
$\tilde{A}_{\parallel} \propto \cos(k_y y)$ with wave number
$k_y a=2\pi a/L_y=0.8$, yielding a value of the tearing instability
parameter $\Delta'a\approx 23.2$.

We solve \revise{the} fully electromagnetic gyrokinetic equations for
electrons and ions using
{\tt AstroGK}~\revise{\cite{NumataHowesTatsuno_10}}. The code employs a
pseudo-spectral algorithm 
for the spatial \revise{coordinates} ($x, y$), and Gaussian quadrature for
velocity space integrals. The velocity space is discretized in the
energy $E_s=m_s v^{2}/2$ ($m_{s}$ is the mass and $s={\mathrm{i}},
{\mathrm{e}}$ is the species label) and
$\lambda=v_{\perp}^{2}/(B_{z0}v^{2})$.

There are four \revise{basic} parameters in the system:
The mass ratio, $\sigma\equiv \me/\mi$, the
temperature ratio of the background plasma, $\tau \equiv \Ti/\Te$, the
electron plasma beta, 
$\betae\equiv n_{0}\Te/(B_{z0}^{2}/2\mu_{0})$
 where $n_0$ is the background plasma density of ions and electrons and 
 $\mu_0$ is the vacuum permeability, and the ratio of the ion sound
 Larmor radius to the equilibrium scale length $a$, $\rhos/a\equiv 
c_{\mathrm{Se}}/(\Omega_{\mathrm{ci}}a)$. The ion sound speed is
$c_{\mathrm{Se}}=\sqrt{\Te/\mi}$, and the ion cyclotron frequency is
$\Omega_{\mathrm{ci}}=e B_{z0}/\mi$. Those parameters define the
physical scales associated with the non-magnetohydrodynamic (MHD)
effects: 
\begin{equation}
 \label{eq:ion_e_scales}
\begin{split}
 \rhoi = \tau^{1/2} \rhos\sqrt{2}, ~~~ 
 \di = \betae^{-1/2} \rhos\sqrt{2}, ~~~ 
 \rhoe = \sigma^{1/2} \rhos\sqrt{2}, ~~~ 
 \de = \betae^{-1/2} \sigma^{1/2} \rhos\sqrt{2}.
\end{split}
\end{equation}

Collisions are modeled by the linearized, and gyro-averaged Landau
collision operator in \revise{{\tt
AstroGK}
\cite{AbelBarnesCowley_08,BarnesAbelDorland_09}}.
There are like-particle collisions of electrons and ions whose
frequencies are given by $\nuee$ and $\nuii$, and inter-species
collisions of electrons with ions given by $\nuei$, which is equal to
$\nuee$ for the current parameters.
The ion-electron collisions are subdominant compared with the ion-ion
collisions. The electron-ion collisions reproduce Spitzer resistivity
for which the electron-ion collision frequency and the resistivity
($\eta$) are related by $\eta/\mu_{0}=0.380 \nuei \de^{2}$. In terms of
the Lundquist number,
$S=\mu_{0}a\VA/\eta=2.63(\nuei\tauA)^{-1}(\de/a)^{-2}$, where $\VA\equiv
B_{y}^{\mathrm{max}}/\sqrt{\mu_{0}n_{0}\mi}$ is
the Alfv\'en velocity corresponding to the peak value of
$B_{y}^{\mathrm{eq}}$, and the Alfv\'en time $\tauA \equiv a/\VA$. We
choose the electron collision frequency $\nue=\nuee=\nuei$ \revise{to be
sufficiently} small so that macroscopic dynamics (e.g., the tearing mode
growth rate) is independent of $\nue$
\referee{(i.e., the frozen flux condition is broken by electron inertia,
not collisions.)}.

\section{Simulation results}

\subsection{Low-$\beta$ case}

As a reference case, we perform a simulation for low-$\betae$ plasmas
($\betae\sim \me/\mi$). In this case, the ion response is essentially
electrostatic~\cite{ZoccoSchekochihin_11}, and only electron heating
matters~\cite{LoureiroSchekochihinZocco_13}. We take
$\beta_{\mathrm{e}}=\sigma=0.01$, $\sqrt{2}\rhos/a = 0.25$, $\tau=1$
($\betai=\betae$), yielding 
$\rhoi/a = \de/a = 0.1 \di/a = 10 \rhoe/a=0.25$.
Collision frequencies are $\nuee\tauA=\nuii\tauA=8\times10^{-5}$
($S\sim5\times10^{5}$).
The spatial resolutions in the $x$, $y$ directions are $N_{x}=256$,
$N_{y}=128$ subject to the $2/3$ rule for dealiasing. The number of the
velocity space collocation points $N_{\lambda}=N_{E}=64$ are determined
to resolve fine structures in velocity space~\cite{NumataLoureiro_12}.

To estimate plasma heating, we measure the collisional energy
dissipation rate,
\begin{equation}
 D_{s} =
  -\int \int
  \left\langle
   \frac{T_{0s}h_{s}}{f_{0s}} 
   \left(\frac{\p h_{s}}{\p t}\right)_{\mathrm{coll}}
  \right\rangle_{\bm{r}}
  \diff \bm{r} \diff \bm{v} \geq 0.
  \label{eq:dissipation_rate}
\end{equation}
Without collisions, the gyrokinetic equation conserves the generalized
energy consisting of the particle part $E^{\mathrm{p}}_{s}$ and the
magnetic field part $E^{\mathrm{m}}_{\perp,\parallel}$
\begin{equation}
 W = \sum_{s} E^{\mathrm{p}}_s + E^{\mathrm{m}}_{\perp} +
  E^{\mathrm{m}}_{\parallel} = 
  \int
  \left[
   \sum_{s}\int \frac{T_{0s} \delta f_{s}^{2}}{2f_{0s}} \diff \bm{v}
   + \frac{\left|\nabla_{\perp}A_{\parallel}\right|^{2}}{2\mu_{0}}
   + \frac{\left|\delta B_{\parallel}\right|^{2}}{2\mu_{0}}
  \right] \diff \bm{r}
\end{equation} 
 where \revise{$\delta f_{s}= - \left(q_{s}\phi/T_{0s}\right) f_{0s} +
 h_{s}$} is the perturbation of the distribution function, $h_{s}$ is the
 non-Boltzmann part obeying the gyrokinetic equation, and the
 generalized energy is dissipated by collisions as $\diff W/\diff t =
 -\sum_{s} D_{s}$. The collisional dissipation increases the entropy
 (related to the first term of the generalized energy), and is turned
 into heat~\cite{HowesCowleyDorland_06}.

\begin{wrapfigure}{r}{60mm}
 \vspace*{-2em}
 \begin{center}
  \includegraphics[scale=0.45,bb=10 -1 410 482]{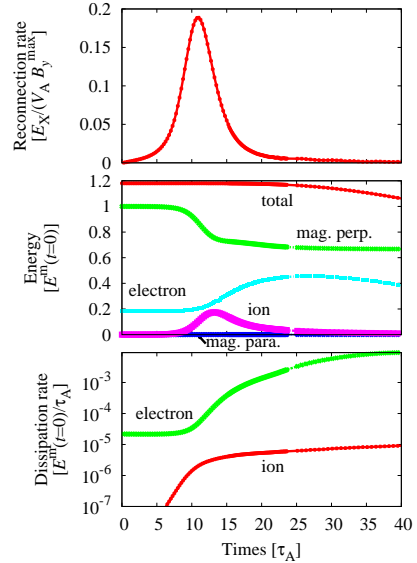}
  \caption{\label{fig:tevo_lowb}Time evolution of the reconnection rate
  (top), the energy components (middle), and the dissipation rate of ions
  and electrons (bottom).}
 \end{center}
\end{wrapfigure}
Figure~\ref{fig:tevo_lowb} shows time evolutions of the reconnection
rate measured by the electric field at the $X$ point
$(x,y)=(L_x/2,L_y/2)$, the energy components normalized by the initial
magnetic energy, and the collisional energy dissipation rate. The peak
reconnection rate \revise{is fast $E_{\mathrm{X}}/(\VA B_{y}^{\mathrm{max}})
\sim 0.2$.}
During magnetic reconnection
the magnetic energy is converted to the particle's energies reversibly. 
\referee{First, the ion $\bm{E}\times\bm{B}$ drift flow is excited, thus
the ion energy increases. Then, electrons exchange energies with the
excited fields through phase mixing process. Electrons store the
increased energy in the form of temperature fluctuations and higher
order moments.}
Collisionally dissipated energy is about 1\% of the initial magnetic
energy after dynamical process has almost ended ($t/\tauA=25$).
The energy dissipation starts to grow rapidly when the maximum
reconnection rate is achieved. It stays large long after the dynamical
stage, and an appreciable amount of the energy is lost in the later
time. As has previously been reported
in~\cite{LoureiroSchekochihinZocco_13,NumataLoureiro_12}, the 
collisional dissipation is independent of the collision frequency in the
collisionless regime indicating phase mixing.
The ion dissipation is negligibly small compared with the electron's
for the low-$\betae$ case.

\begin{figure}[htbp]
 \begin{center}
 \includegraphics[scale=0.5,bb=10 0 306 271]{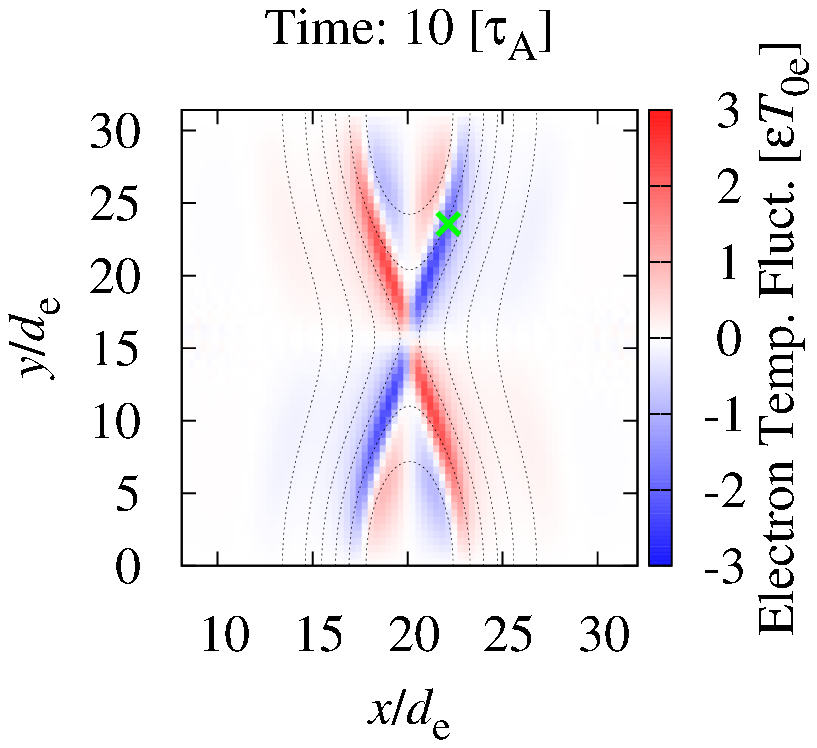}
 \includegraphics[scale=0.5,bb=10 0 450 302]{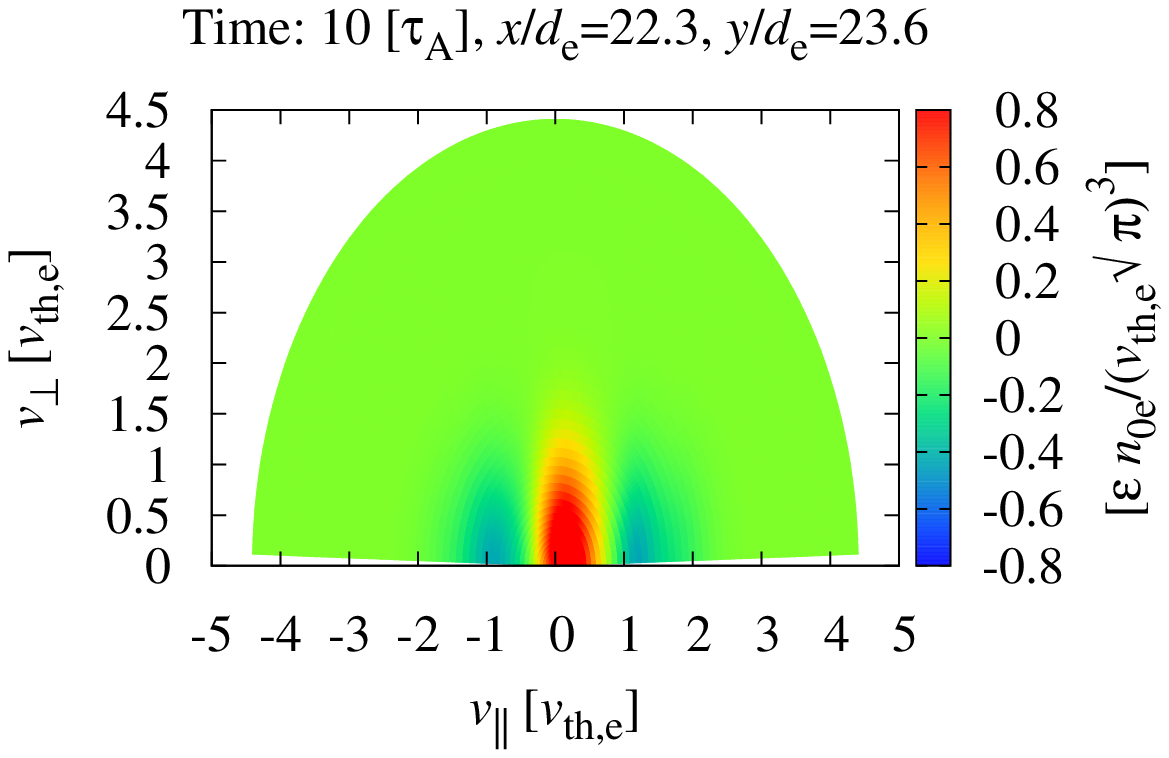}\\
 \includegraphics[scale=0.5,bb=10 0 306 271]{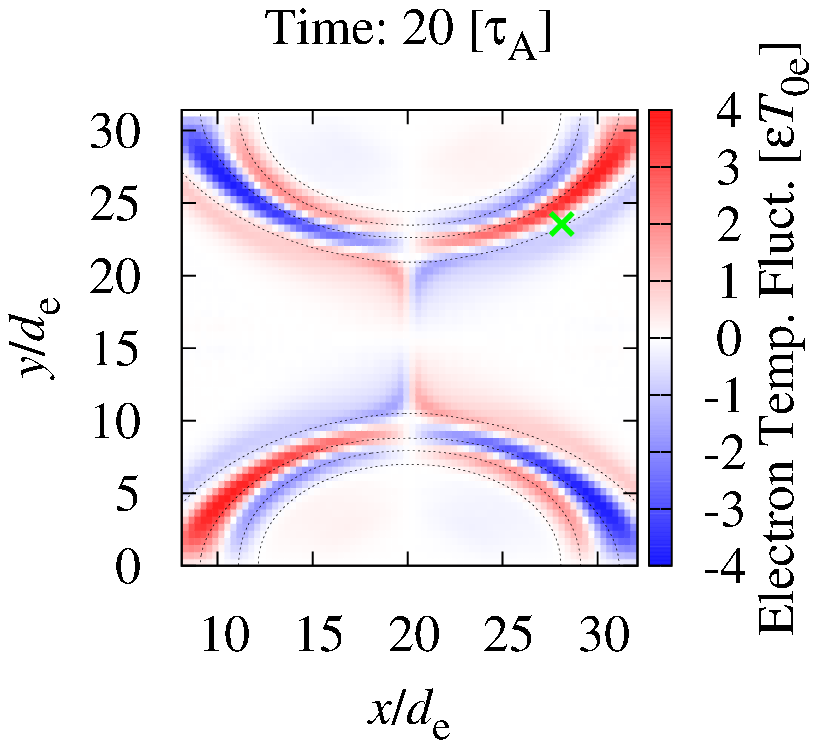}
 \includegraphics[scale=0.5,bb=10 0 450 302]{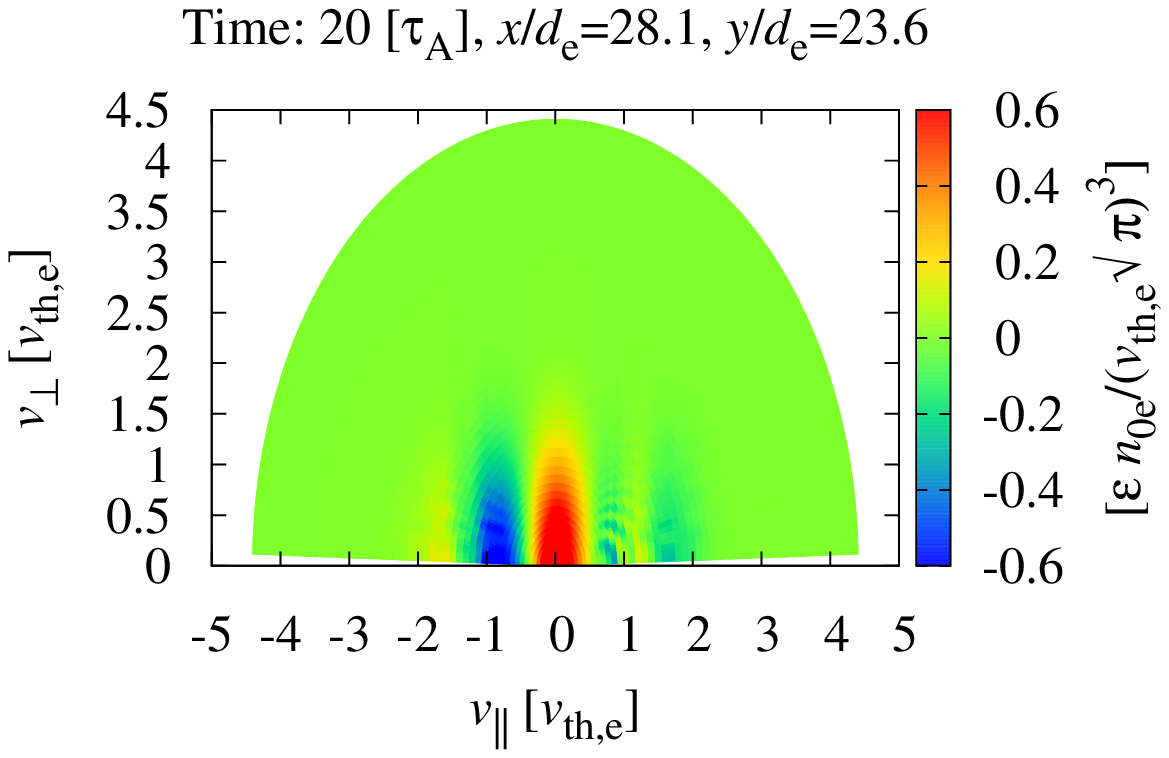}
 \caption{\label{fig:vstruct}Parallel electron temperature fluctuations
  and velocity space structures of the electron distribution function.
  Isolines of the magnetic flux are over-plotted with the temperature
  fluctuation. 
  \referee{The distribution functions are taken near the separatrix.}}
 \end{center}
\end{figure}

To illustrate the phase mixing process, we show the parallel electron
temperature fluctuation and electron 
distribution function in velocity space in Fig.~\ref{fig:vstruct} at
$t/\tauA=10, 20$.
\referee{We subtract the lower order moments corresponding to the
density and parallel flow from $h$ to clearly see phase-mixing structures.}
The distribution functions are taken
\referee{near the separatrix denoted by the cross marks in the left panels.}
In the earlier time of the nonlinear stage ($t/\tauA=10$), the
distribution function only has gradients in the $v_{\parallel}$
direction indicating parallel phase mixing. We note that since there is
no variation in the $z$ direction, parallel phase mixing occurs along
the perturbed field lines.
\referee{Later, the perpendicular FLR phase mixing follows to create structures
in the $v_{\perp}$ direction although the effect seems weak because
$k_{\perp}\rhoe$ is small.}
The occurrence of the perpendicular phase mixing highlights the
difference from~\cite{LoureiroSchekochihinZocco_13}. About 10\% of the
initial magnetic energy, \referee{which accounts for $\sim36$\% of the
released magnetic energy,} is dissipated \revise{at $25\leq t/\tauA\leq40$}.
\referee{The rest of the released magnetic energy is used to excite
electron temperature perturbations.}

\subsection{High-$\beta$ case}

\begin{wrapfigure}{r}{60mm}
 \vspace{-2em}
 \begin{center}
  \includegraphics[scale=0.45,bb=10 10 410 302]{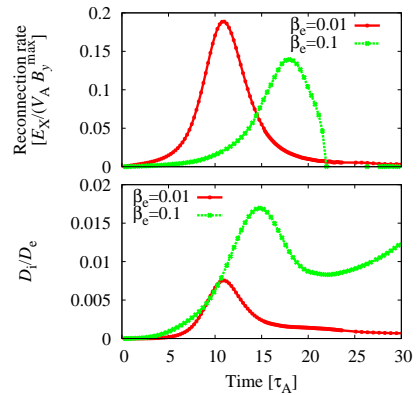}
  \caption{\label{fig:hrate}Reconnection rate and ratio of ion to
  electron dissipation rate $D_{\mathrm{i}}/D_{\mathrm{e}}$ for
  $\betae=0.01, 0.1$.}
 \end{center}
\end{wrapfigure}

In high-$\beta$ plasmas, compressible fluctuations will be excited which
are strongly damped collisionlessly \revise{(see, e.g., Sec.~6 of
\cite{SchekochihinCowleyDorland_09})}. This may open up another 
dissipation channel where phase mixing of the ion distribution function
ends up \revise{in} ion heating.
We compare the ratio of the energy dissipation rate for ions and
electrons $D_{\mathrm{i}}/D_{\mathrm{e}}$ for $\betae=0.01, 0.1$ in
Fig.~\ref{fig:hrate}.
\revise{For $\betae=0.01$ case,} the ion dissipation rises prior to
the electron dissipation, and the heating ratio peaks at about the same
time as the peak reconnection rate.
\revise{For the high-$\betae$ case, the reconnection rate is peaked
around $t/\tauA=18$ because linear growth is slower, and it sharply
drops around $t/\tauA=22$ where a small island like structure is 
generated at the $X$-point.}
\revise{$D_{\mathrm{i}}/D_{\mathrm{e}}$ for $\betae=0.1$ peaks before
the maximum reconnection rate, and reaches twice as high as that for
$\betae=0.01$. The ratio increases again from $t/\tauA=20$ because the
electron dissipation starts to decrease.}
 As anticipated, the heating ratio of ions to electrons
increases with increasing $\betae$ though it is still small.
The ion heating may be relevant only for much higher-$\betae$ plasmas.

\section{Conclusions}

We have performed gyrokinetic simulations to investigate plasma heating
during magnetic reconnection in strongly magnetized, weakly collisional
plasmas. In the collisionless limit where the macroscopic behavior of
plasmas is insensitive to collisions, parallel and perpendicular phase
mixing create fine structures in velocity space leading to strong energy
dissipation. We have shown \revise{that a} significant amount 
of the initial magnetic energy is converted into electron heating for
low-$\betae$ plasmas. The electron heating occurs after the dynamic
reconnection process ceased, and continues for longer
time. 
\revise{We observe perpendicular phase mixing as well as parallel phase mixing
in the nonlinear regime.}
The low-$\betae$ case result is consistent with the
previous study using the hybrid fluid/kinetic model
in~\cite{LoureiroSchekochihinZocco_13} except for the perpendicular
phase mixing.

We have also shown \revise{a relatively high-$\betae$ case}. For
$\betae=\betai=0.1$, the ion heating rate becomes larger compared with
the $\betae=\betai=0.01$ case although it is still small compared
with that of electrons. The ion heating may be important only for much
higher-$\beta$ plasmas.

\section*{Acknowledgments}

This work was supported by JSPS KAKENHI Grant Number 24740373. This work
was carried out using the HELIOS supercomputer system at Computational
Simulation Centre of International Fusion Energy Research Centre
(IFERC-CSC), Aomori, Japan, under the Broader Approach collaboration
between Euratom and Japan, implemented by Fusion for Energy and JAEA.
\revise{NFL was supported by Funda\c{c}\~ao para a Ci\^encia e Tecnologia
(Ci\^encia 2008 and Grant No. PTDC/FIS/118187/2010) and by the European
Communities under the Contracts of Association between EURATOM and IST.}

\bibliographystyle{jpsj}
\providecommand{\noopsort}[1]{}

\end{document}